\documentclass[aps,prd,preprint,superscriptaddress]{revtex4}
\usepackage{graphicx}
\usepackage{amsmath}
\usepackage{bm}% bold math
\usepackage{color}% bold math
\usepackage{amssymb,enumerate}

\def\ee{\end{eqnarray}}

\def\=:{=\hspace{-.7em}\raisebox{1.1ex}{.}\hspace{.1em}\raisebox{-0.2ex}{.} }
\def\ee{\end{eqnarray}}

\def\=:{=\hspace{-.7em}\raisebox{1.1ex}{.}\hspace{.1em}\raisebox{-0.2ex}{.} }

\newcommand {\beq}{\begin{eqnarray}}
\newcommand {\eeq}{\end{eqnarray}}
\newcommand {\non}{\nonumber\\}

\newcommand {\1}[1]{\frac{1}{#1}}

\newcommand {\ph}{\varphi}

\newcommand {\del}{\partial}

\newcommand {\tr}{{\rm tr}\,}

\begin{document}

%\preprint{}

\title{Josephson junction of non-Abelian superconductors \\
and non-Abelian Josephson vortices
}


\author{Muneto Nitta}

%\thanks{}
\affiliation{
Department of Physics, and Research and Education Center for Natural 
Sciences, Keio University, Hiyoshi 4-1-1, Yokohama, Kanagawa 223-8521, Japan\\
}


\date{\today}
\begin{abstract}
A Josephson junction is made of two superconductors sandwiching 
an insulator, and a Josephson vortex is a magnetic vortex 
(flux tube) absorbed into the Josephson junction, whose dynamics can be 
described by the sine-Gordon equation. 
In a field theory framework, 
a flexible Josephson junction was proposed,
in which the Josephson junction is represented by 
a domain wall separating two condensations 
and a Josephson vortex is a sine-Gordon soliton 
in the domain wall effective theory. 
In this paper, we propose a Josephson junction of 
non-Abelian color superconductors, that is 
described by a non-Abelian domain wall, 
and show 
that a non-Abelian vortex (color magnetic flux tube) absorbed into it 
is a non-Abelian Josephson vortex represented as 
a non-Abelian sine-Gordon soliton 
in the domain wall effective theory,
that is the $U(N)$ principal chiral model.

\end{abstract}
%\keywords{}

\maketitle

\section{Introduction}

Superconductivity is one of the most important phenomena
in condensed matter physics.
A Josephson junction is made of 
an insulator sandwiched by 
two superconductors with condensates 
$\Psi_1$ and $\Psi_2$, 
in which the tunneling effect introduces a 
Josephson term $\Psi_1^*\Psi_2$.
When a magnetic field is applied to a type-II superconductor
the magnetic flux is squeezed into vortices.
In the case of a Josephson junction of two 
type-II superconductors, 
 vortices in the bulk are absorbed into the insulator, 
becoming Josephson vortices or fluxons; 
see Ref.~\cite{Ustinov:1998} as a review. 
It is known that dynamics of Josephson vortices 
can be described by the sine-Gordon equation. 
Josephson vortices also appear in high-$T_{\rm c}$ superconductors 
with multi-layered structures \cite{Blatter:1994} and in 
two coupled Bose-Einstein condensates \cite{Kaurov:2005}.
It is also known that 
not only Josephson junctions 
but also multi-band superconductors 
have intrinsic Josephson terms 
to allow sine-Gordon solitons 
\cite{Tanaka:2001,Gurevich:2003,Goryo:2007,Tanaka:2015}. 

In a field theory framework, 
a flexible Josephson junction was proposed in Ref.~\cite{Nitta:2012xq},
in which the Josephson junction is represented by 
a domain wall reducing to the usual Josephson junction 
in the heavy domain wall limit.
Vortices in the bulk become sine-Gordon kinks inside the domain wall 
\cite{Nitta:2012xq,Kobayashi:2013ju}
as usual Josephson junctions. 
In the strong gauge coupling limit, the model 
is reduced to the ${\mathbb C}P^1$ model, 
and the domain wall is reduced to a ${\mathbb C}P^1$ domain wall 
\cite{Abraham:1992vb, Arai:2002xa} 
which is also magnetic domain wall in ferromagnets, 
e.g.~\cite{Kobayashi:2014xua}.
The Josephson term introduces 
the sine-Gordon potential 
in the effective theory of the $d=1+1$ dimensional 
domain wall world-volume, 
and a sine-Gordon solitons carries a quantized magnetic flux, 
corresponding to a magnetic vortex in the bulk 
superconductors.  
This correspondence has been generalized to 
higher dimensional Skyrmions 
\cite{Nitta:2012wi} 
and to Yang-Mills instantons 
\cite{Nitta:2013cn,Nitta:2013vaa}.

In this paper, we discuss a Josephson junction of 
non-Abelian color superconductors and 
non-Abelian Josephson vortices in it. 
Non-Abelian superconductors can be described by
a $U(N)$ (or $SU(N)$) gauge theory with $N$ scalar fields in the fundamental 
representation. 
$U(N)$ gauge symmetry is spontaneously broken, 
and therefore is referred to as a non-Abelian or color superconductor.
One example is supersymmetric gauge theories 
in the Higgs phase, studied extensively in these years 
\cite{Tong:2005un,Eto:2006pg,Shifman:2007ce}.
The other example is the color-flavor locked phase of 
QCD at extremely high density \cite{Alford:2007xm,Eto:2013hoa}.
We discuss physics that appears 
when two such non-Abelian superconductors 
are connected as a non-Abelian Josephson junction 
that is flexible.
Instead of 
the domain wall of the above $U(1)$ superconductors, 
a non-Abelian Josephson junction can be represented by 
a non-Abelian domain wall 
\cite{Shifman:2003uh,Eto:2005cc,Eto:2008dm}. 
On the other hand, vortices in non-Abelian superconductors 
are non-Abelian vortices or color magnetic flux tubes. 
Non-Abelian vortices were first found in 
supersymmetric theories \cite{Hanany:2003hp,Auzzi:2003fs,
Shifman:2004dr,Eto:2005yh}, 
and they carry non-Abelian ${\mathbb C}P^{N-1}$ moduli;
see Refs.~\cite{Tong:2005un,Eto:2006pg,Shifman:2007ce} for 
a review.
Non-Abelian vortices in high density QCD were 
found in Ref.~\cite{Balachandran:2005ev}
and they also carry   
 ${\mathbb C}P^2$ moduli \cite{Nakano:2007dr};
see Ref.~\cite{Eto:2013hoa} for a review. 
Therefore, when a non-Abelian vortex is placed parallel to 
a non-Abelian vortex, the former is absorbed into the latter 
to minimize the energy as parallel to usual Josephson junctions. 
In order to find what the fate of this non-Abelian vortex,
we consider the effective theory approach. 
The effective field theory of a non-Abelian domain wall 
is the $U(N)$ chiral Lagrangian or principal chiral model 
\cite{Shifman:2003uh,Eto:2005cc,Eto:2008dm}. 
We show that conventional (quadratic) non-Abelian Josephson 
term introduced in the model induces 
the conventional (modified or quadratic) pion mass term in 
the domain wall effective theory. 
This model has been recently found to admit 
a non-Abelian sine-Gordon soliton \cite{Nitta:2014rxa}
that carries ${\mathbb C}P^{N-1}$ moduli \cite{Eto:2015uqa}.
By calculating the non-Abelian color magnetic flux, 
we find that the non-Abelian sine-Gordon soliton is precisely 
the non-Abelian vortex that is absorbed into the non-Abelian 
Josephson junction (domain wall), 
so we call it non-Abelian Josephson vortex. 
For the quadratic Josephson term, we find that 
a non-Abelian sine-Gordon kink carries a half color magnetic flux of 
the non-Abelian vortex in the bulk. 
Therefore, one non-Abelian vortex is split into two 
color flux tubes inside the domain wall.
We also discuss 3+1 dimensional configurations of Josephson vortices, 
suggesting a non-Abelian extension of a D-brane soliton.

This paper is organized as follows. 
After our model is given in Sec.~\ref{sec:model}, 
we present the main results in Sec.~\ref{sec:SGkink}; 
in the presence of a non-Abelian domain wall, 
we construct the domain wall effective theory by 
the moduli approximation to obtain 
the $U(N)$ chiral Lagrangian. 
We add the non-Abelian Josephson term in the original theory 
and find that it induces a pion mass term in 
the effective theory. 
We then construct a non-Abelian sine-Gordon soliton 
on the domain wall 
and show that it carries a non-Abelian magnetic flux, 
to show the coincidence between  
the non-Abelian sine-Gordon soliton 
and the non-Abelian vortex. 
Section \ref{sec:summary} is devoted to a summary 
and discussion. 

\newpage

%%%%%%%%%%%%%%%%%%%%%%%%%%
\section{The model \label{sec:model}}
We consider the $U(N)$ gauge theory 
coupled with two $N$ by $N$ charged complex scalar fields 
$H_1(x)$ and $H_2(x)$ summarized by $H = (H_1,H_2)$, 
with massless and real $N$ by $N$ scalar field $\Sigma(x)$ in $d=2+1$ dimensions. 
The Lagrangian that we consider is given by
\beq
&& {\cal L} = -\1{4 g^2} \tr F_{\mu\nu}F^{\mu\nu} 
 + \1{g_2^2} \tr (D_{\mu} \Sigma)^2 
 + \tr |D_{\mu} H|^2 - V\\
&& V = {\lambda \over 4} \tr (H H^\dagger  -v^2 {\bf 1}_N)^2 
 + \tr |\Sigma H - H M|^2 
\eeq
where the covariant derivatives are 
$D_{\mu} H = \del_{\mu} H - i A_{\mu } H$ 
and 
$D_{\mu} \Sigma = \del_{\mu} \Sigma - i [A_{\mu},\Sigma]$, 
$g$ is the gauge coupling constant common for 
$U(1)$ and $SU(N)$ factors, 
$g_2$ and $\lambda$ are coupling constants, 
and the masses of $H$ are given by 
$M={\rm diag.}(m_1 {\bf 1}_N,m_2 {\bf 1}_N)$ with $m_1>m_2$.
The symmetry of the model is
$U(N)$ gauge (color) symmetry and global (flavor) symmetries
\beq
 A_{\mu} \to g A_{\mu} g^{-1} + i g \del_{\mu} g^{-1},  \quad
 \Sigma \to g \Sigma g^{-1}, \quad
 H \to g H , \quad g \in U(N)_{\rm C}\\
 H_1 \to H_1 U_{\rm L}, \quad 
 H_2 \to H_2 U_{\rm R},  \quad U_{\rm L,R} \in SU(N)_{\rm L,R}.
\label{eq:flavor}
\eeq

In the limit $g_2=g$, $\lambda/2=g^2$, the model enjoys 
${\cal N}=4$ supersymmetry (with eight supercharges) 
in $d=2+1$ by doubling
scalar fields $H$ and adding 
fermion superpartners; see, e.g., Ref.~\cite{Eto:2006pg} for a review. 
In this paper, supersymmetry is not essential apart from technical reasons.

When $g_2 \to \infty$, 
the kinetic term of $\Sigma$ disappears, 
then it becomes 
an auxiliary field that can be eliminated by its equation of motion as
\beq
 \Sigma = {H M H^\dagger \over H H^\dagger}.
\eeq
If we further take $\lambda \to \infty$, 
the above Lagrangian is reduced to
\beq
&& {\cal L} = -\1{4 g^2} \tr F_{\mu\nu}F^{\mu\nu} 
 + \tr |D_{\mu} H|^2 - V\\
&& V =  
 -\tr [ v^{-2} (H M H^\dagger)^2  + H M^2 H^\dagger] 
 = {4 m^2 \over v^2} \tr (H_1H_1^\dagger H_2H_2^\dagger).
\eeq

Without taking any limits, we may consider 
the Lagrangian 
\beq
&& {\cal L}_2 = -\1{4 g^2} \tr F_{\mu\nu}F^{\mu\nu} 
 + \tr |D_{\mu} H|^2 - V_2\\
&& V_2 
= {\lambda \over 4} \tr (H H^\dagger -v^2  {\bf 1}_N)^2 
 + {4 m^2 \over v^2} \tr (H_1H_1^\dagger H_2H_2^\dagger)
\eeq
from the beginning, instead of the above procedure.

Let us discuss the vacuum structure of the model.
In the massless case $m_1=m_2=0$, 
the flavor symmetry is enhanced to $SU(2N)$, 
and 
the vacuum can be taken to be
\beq
 H 
= 
\left(
 v {\bf 1}_N ,{\bf 0}_N 
\right) ,
\quad
\quad \Sigma ={\bf 0}_N
\eeq 
by using the $SU(2N)$ flavor symmetry.
This is the so-called color-flavor locked vacuum.
The moduli space of vacua is 
the Grassmann manifold, see, e.g., Ref.~\cite{Higashijima:1999ki}:
\beq 
Gr_{2N,N}\simeq {SU(2N) \over SU(N)\times SU(N)\times U(1)}.
\label{eq:Grassmann}
\eeq 
In the massive case, $m_1,m_2 \neq 0$, 
the vacuum is split into disjoint vacua 
\beq
H = 
\left(
 v {\bf 1}_N ,{\bf 0}_N 
\right) , \quad \Sigma =m_1 {\bf 1}_N,
\quad
{\rm or}
\quad 
H = \left(
  {\bf 0}_N , v{\bf 1}_N 
\right) ,  \quad \Sigma =m_2 {\bf 1}_N,
 \label{eq:vac}
\eeq 
with the following unbroken {\it global} symmetries, respectively \cite{footnote1}: 
\beq
SU(N)_{\rm C+L} ,
\quad 
 {\rm or} \quad 
 SU(N)_{\rm C+R}.
\eeq
Each vacuum given here can be interpreted as 
a non-Abelian superconductor, 
since gauge group $U(N)$ is spontaneously broken.

For explicit calculation, 
we work in the strong gauge coupling limit $g^2 \to \infty$ 
in which the gauge field becomes non-dynamical and
can be eliminated by its equation of motions as
\beq 
 A_{\mu} 
= {i\over 2} v^{-2} [H \del_{\mu} H^\dagger -  (\del_{\mu}H)  H^\dagger].
\eeq
The model reduces to a Grassmann sigma model 
with the target space in Eq.~(\ref{eq:Grassmann})
and a potential term, 
known as the massive Grassmann model \cite{Arai:2003tc}.
We denote the limit $\lambda/2=g^2=g_2^2 \to \infty$ 
the sigma model limit.
Although we take this limit for explicit calculation,
the results in this paper do not rely on this limit.

As we will see in the next subsection, 
the above model admits a domain wall solution 
interpolating the two vacua in Eq.~(\ref{eq:vac}), 
that separate the condensation $H_1$ and $H_2$. 
In order to interpret this domain wall as a Josephson junction, 
we now introduce a deformation 
\beq 
{\cal L}_{J,1} = -
 \gamma \tr (H_1^\dagger H_2  + H_2^\dagger H_1) 
\label{eq:Josephson}
\eeq 
that explicitly breaks the flavor symmetry in Eq.~(\ref{eq:flavor}) 
to $SU(N)_{\rm L+R}$. 
We refer this term the non-Abelian ``Josephson" interaction term 
\cite{Nitta:2013cn,Nitta:2014rxa}, 
because it is a non-Abelian matrix extension of a Josephson term in 
the Josephson junction of two superconductors 
with two condensations. 
In the case of supersymmetric extension of the model,  
the Josephson term in Eq.~(\ref{eq:Josephson}) 
breaks supersymmetry explicitly 
but supersymmetry is not essential in our study.
Instead of the Josephson term in Eq.~(\ref{eq:Josephson}),
we may also consider a quadratic Josephson-like term
\beq 
{\cal L}_{J,2} = -
 \gamma \tr [(H_1^\dagger H_2)^2  + (H_2^\dagger H_1)^2 ].
\label{eq:Josephson2}
\eeq 
This is a non-Abelian extension of the Josephson-like 
term in chiral p-wave superconductors \cite{Agterberg1998,Garaud:2012}. 
We refer it the non-Abelian quadratic Josephson term.

%%%%%%%%%%%%%%%%%%%%%%%%%%%%%%%%%%%%%%%%%%%%%%%%%%
\section{Non-Abelian Josephson vortex: non-Abelian sine-Gordon soliton inside a 
non-Abelian domain wall
\label{sec:SGkink}}

\subsection{Non-Abelian domain wall}
Meanwhile we consider the case 
in the absence of the 
Josephson term $\gamma=0$, 
and we turn it on later. 
A non-Abelian domain wall solution 
interpolating between the left and right vacua, 
which is placed perpendicular to the $x^2$ coordinate, 
is given by \cite{Isozumi:2004jc,
Shifman:2003uh,Eto:2005cc,Eto:2008dm}
\beq
&& H_{\rm wall,0} = {v \over \sqrt {1+|u_{\rm wall}|^2}}
      \left({\bf 1}_N, u_{\rm wall} {\bf 1}_N\right),
\quad
u_{\rm wall}(x^2) = e^{\mp m (x^2-Y) + i \ph} , 
  \label{eq:wall-sol0}\\
&& \Sigma_{\rm wall,0} = v^{-2}H M H^\dagger ,
\quad 
A_{2,{\rm wall},0} 
= {i\over 2} v^{-2} [H\del_2 H^\dagger - (\del_2 H)  H^\dagger],
\nonumber
\eeq
in the sigma model limit. 
The most general solution can be obtained by 
acting the $SU(N)_{\rm C+L+R}$ symmetry on the particular solution in Eq.~(\ref{eq:wall-sol0}):
\beq 
&& H_{\rm wall} = V H_{\rm wall,0} 
\left(\begin{array}{cc}
 V^\dagger & 0 \\ 0 & V
\end{array} \right) 
=  {v\over \sqrt {1+ e^{\mp 2 m (x^2-Y) }}}
      \left({\bf 1}_N, e^{\mp m (x^2-Y) }U \right), 
\non
&&
 \Sigma_{\rm wall} = V \Sigma_{\rm wall,0} V^\dagger ,
\quad
A_{2,{\rm wall}} = V A_{2,{\rm wall},0} V^\dagger ,
\eeq 
with $V \in SU(N)$ and 
$U \equiv V^2 e^{i\ph} \in U(N)$.
Therefore, the domain wall has the moduli \cite{footnote2}
\beq 
 {\cal M}_{\rm wall} \simeq {\mathbb R} \times U(N).\label{eq:wall-moduli}
\eeq 
In the presence of the Josephson term $\gamma \neq 0$,
this domain wall behaves as a Josephson junction 
as we will see below.

%%%%%%%%%%%%%%
\subsection{Non-Abelian vortex}

Let us discuss a non-Abelian vortex.
In the left vacuum in Eq.~(\ref{eq:vac}), 
we can neglect $H_2$.
There, $U(N)$ symmetry is spontaneously broken 
and is locked with the $SU(N)_{\rm L}$ flavor symmetry 
to be the $SU(N)_{\rm C+L}$ color-flavor locked symmetry.
There is a non-Abelian vortex solution 
with a non-Abelian magnetic field and a scalar field given by 
\beq
 && F_0 = \, {\rm diag} (F_*(r),0,\cdots,0), \quad
 H_0 = v \, {\rm diag} (f(r)e^{i \theta},1,\cdots,1),
 \label{eq:NA-vortex}
\eeq
respectively, with the boundary conditions 
$f(r) \to 1$ ($r\to \infty$) and 
$f(r) \to 0$ ($r=0$),
where $(r,\theta)$ are cylindrical coordinates.  
This solution is typically obtained by  
embedding of the Abrikosov-Nielsen-Olesen (ANO) 
vortex solution \cite{Abrikosov:1956sx} 
$(F_*(r),f(r)e^{i \theta})$ in the upper-left corner. 
The integral of the magnetic field is quantized:
\beq
 \int d^2x F_0 = {\rm diag} (2\pi,0,\cdots,0) .
\eeq

The most general solution is obtained by acting the 
color-flavor locked symmetry $SU(N)_{\rm C+L}$: 
\beq
 && F = V {\rm diag} (F_*(r),0,\cdots,0) V^\dagger, \quad
  H = v \, V{\rm diag} (f(r)e^{i \theta},1,\cdots,1) V^\dagger, \non
&& V \in SU(N).
 \label{eq:NA-vortex}
\eeq
In other words,
the solution spontaneously breaks 
the color-flavor locked symmetry $SU(N)_{\rm C+L}$ 
into a subgroup 
$SU(N-1)\times U(1)$, and consequently 
there appears moduli localized on the vortex core;
\beq
{\cal M}_{\rm vortex} 
\simeq   {\mathbb C} \times {\mathbb C}P^{N-1} 
= {\mathbb C} \times {SU(N)_{\rm C+L} \over SU(N-1)\times U(1)}.
\label{eq:vortex-moduli}
\eeq

We can repeat the same for the right vacuum. 
The Josephson term does not affect 
the non-Abelian vortex living in each vacuum. 
When scalar fields $H$ are massless, 
vortices are so-called non-Abelian semi-local vortices 
\cite{Shifman:2006kd}, 
which reduce to Grassmann lumps in the sigma model limit.
In the presence of the mass term as we are discussing here, 
vortices are local vortices.

If we consider a non-Abelian vortex parallel to a non-Abelian vortex, 
the latter will be absorbed into the former  
to minimize the total energy, in analogy with usual superconductors. 
In this case, the Josephson term plays an essential role.
A question is what the fate of the vortex if it is absorbed 
into the wall.

One comment is in order here.
In the following, we will take the sigma model limit 
for concrete calculations.
In this limit, the bulk Grassmann lumps 
become singular (a delta function) 
in the presence of the mass term,
known as small lump singularity.
Nevertheless, they can live stably  inside the domain wall,  
as we will show below.
The singularity appearing in the bulk lumps is an artifact of 
the sigma model limit, and 
small lumps are replaced by the ANO vortices 
without taking that limit.

%%%%%%%%%%%%%%%%%%%%%%%%%%%%%%%%%%%
\subsection{Low-energy effective theory on non-Abelian domain wall world-volume}

By using the moduli approximation \cite{Manton:1981mp,Eto:2006uw},
the effective theory of the domain wall can be constructed  
by promoting the moduli $X$ and $U$ to fields 
 $X(x^i)$ and $U(x^i)$, respectively ($i=0,1$) 
on the world volume of the domain wall. 
The result is \cite{Shifman:2003uh,Eto:2005cc,Eto:2008dm}:
\beq
 {\cal L}_{\rm wall} =  - {v^2 \over 4m} 
\tr \left(U^\dagger \del_{i} U
            U^\dagger \del^{i} U \right) 
+ {v^2 \over 2m} \del_i X \del^i X .\label{eq:eff}
\eeq
Apart from the position modulus $X$, 
this is the $U(N)$ chiral Lagrangian or principal chiral model.

Here we turn on the Josephson term perturbatively in 
the regime $\gamma << m$, 
and consider the effect on the domain wall effective action. 
We thus obtain 
\beq
 {\cal L}_{{\rm wall},J,1}
&=& - \gamma \int^{+\infty}_{-\infty} dx^2
 {  e^{\mp m (x^2-Y)} \over 1+e^{\mp 2 m (x^2-Y) }} 
 (\tr U + \tr U^\dagger)
= - {\pi \gamma \over 2m} (\tr U + \tr U^\dagger) \non
&=& - m'^2  (\tr U + \tr U^\dagger) ,\quad
\label{eq:eff-Josephson}
m'^2 \equiv  {\pi \gamma \over 2m}
\eeq
This term introduces the conventional pion mass term 
in the $U(N)$ chiral Lagrangian in Eq.~(\ref{eq:eff}).
The $U(N)$ target space is lifted by this potential term, 
leaving the unique vacuum
\beq
 U = {\bf 1}_N .
\eeq

In the presence of the non-Abelian quadratic Josephson term
in Eq.~(\ref{eq:Josephson2}) instead of the linear term, 
the following term is induced in the domain wall effective theory: 
\beq
 {\cal L}_{{\rm wall},J,2}
&=& - \gamma_2 \int^{+\infty}_{-\infty}  dx^2
 \left({  e^{\mp m (x^2-Y)} \over 1+e^{\mp 2 m (x^2-Y) }}\right)^2
 (\tr U^2 + \tr U^{\dagger 2})
= - {\pi \gamma_2 \over 4m} (\tr U^2 + \tr U^{\dagger2}) \non
&=& - m_2'^2  (\tr U^2 + \tr U^{\dagger 2}), \quad
\label{eq:eff-Josephson2}
m_2'^2 \equiv  {\pi \gamma_2 \over 4m}.
\eeq
This mass term is sometimes called a modified pion mass 
in the context of the $SU(2)$ Skyrme model 
\cite{Kudryavtsev:1999zm,Nitta:2012wi}. 
In this case, there are two discrete vacua
\beq
 U = \pm {\bf 1}_N .
\eeq
There exists a domain wall interpolating them \cite{Nitta:2012wi}.

%%%%%%%%%%%%%%%%
\subsection{Non-Abelian sine-Gordon soliton inside 
a non-Abelian domain wall}

The $U(N)$ chiral Lagrangian in Eq.~(\ref{eq:eff}) 
with the mass term in Eq.~(\ref{eq:eff-Josephson}) 
admits non-Abelian sine-Gordon solitons 
\cite{Nitta:2014rxa,Eto:2015uqa}. A single soliton solution is
\beq
&& U(x) = {\rm diag} (u(x^1),1,\cdots,1) ,  \label{eq:embedding}\\
&& u(x^1) = \exp i \theta_{\rm SG}(x^1) 
= \exp \left(4 i \, \arctan \exp [m'' (x^1- X)] \right) ,
\quad m''^2 \equiv  {2\pi\gamma \over v^2}
\label{eq:U(1)-one-kink}
\eeq
The most general solution is given by 
acting the $SU(N)$ symmetry on it:
\beq
 U(x) = V{\rm diag} (u(x^1),1,\cdots,1) V^\dagger , 
  \quad V\in SU(N).  \label{eq:SU(N)moduli}
\eeq
By removing the redundancy, $V$ takes a value in a coset space
\beq
 V \in {SU(N) \over SU(N-1) \times U(1)} \simeq {\mathbb C}P^{N-1},
\eeq
and consequently the moduli space of the sine-Gordon soliton is
found to be 
\beq
{\cal M}_{\rm SG\,soliton} = {\mathbb R} \times {\mathbb C}P^{N-1}.
\label{eq:SG-moduli}
\eeq
The total composite configuration is therefore:
\beq 
H_{\rm composite} 
&=&  \1{\sqrt {1+ e^{\mp 2 m (x^2-Y) }}}
      \left({\bf 1}_N, e^{\mp m (x^2-Y) }V{\rm diag} 
(e^{i \theta_{\rm SG}(x^1)},1,\cdots,1) V^\dagger \right) .\;\;
\label{eq:total}
\eeq
This configuration with the upper sign goes to
\beq
H_{\rm composite}  \to 
\bigg\{
\begin{array}{cc}
  ({\bf 1}_N, {\bf 0}_N), & x^2 \to -\infty \\
  ({\bf 0}_N, {\bf 1}_N) V (e^{i \theta_{\rm SG}(x^1)},1,\cdots,1) V^{\dagger},
 & x^2 \to +\infty
\end{array} \label{eq:total-limit}
\eeq

%%%%%%%%%%%%%%%%%%%%%%
\begin{figure}[ht]
\begin{center}
\includegraphics[width=0.6\linewidth,keepaspectratio]{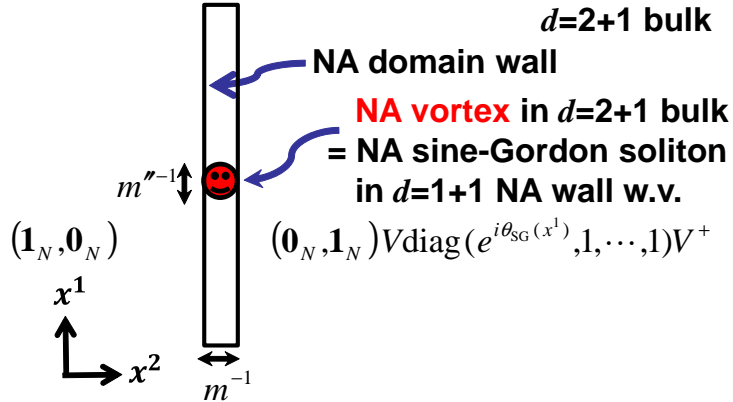}
\caption{A schematic picture of 
a non-Abelian sine-Gordon soliton in 
a non-Abelian domain wall describing 
a non-Abelian vortex. 
\label{fig:SG}}
\end{center}
\end{figure}
%%%%%%%%%%%%%%%%%%%%%%

What is this solution in the $d=2+1$ dimensional  bulk  theory? 
Our claim is that it is precisely a non-Abelian vortex, 
as illustrated in Fig.~\ref{fig:SG}.
The agreement between them 
is not only 
the moduli in Eq.~(\ref{eq:vortex-moduli}) 
for a vortex 
and Eq.~(\ref{eq:SG-moduli}) for a sine-Gordon soliton 
but also the non-Abelian fluxes ($a=1,2$):
\beq
\int d^2 x F_{12} 
 &=& \oint dx^a A_a 
=\oint dx^a {i\over 2} v^{-2} 
[H \del_{\mu} H^\dagger -  (\del_{\mu}H)  H^\dagger]\non
&=& 
V {\rm diag} \left(\int_{-\infty}^{+\infty} dx^1 \del_1 \theta_{\rm SG} |_{x^2=+\infty}
,0,\cdots,0\right) V^\dagger \non
&=& 
V
{\rm diag} \left(
[\theta_{\rm SG} ]_{(x^1,x^2)=(-\infty,+\infty)}^{(x^1,x^2)=(+\infty,+\infty)}  
,0,\cdots,0 \right) V^\dagger\non
&=& V {\rm diag} (2 \pi k, 0,\cdots,0) V^\dagger. \label{eq:flux-matching}
\eeq
Here we have used the sigma model limit in the second equality  
and Eq.~(\ref{eq:total-limit}) for the third equality,
and have assumed 
the $k$ winding  of the phase $\theta_{\rm SG}$ 
for $k$ sine-Gordon solitons in the last equality.   
The flux in Eq.~(\ref{eq:flux-matching}) 
coincides with that of the non-Abelian vortex 
in Eq.~(\ref{eq:NA-vortex})
showing the one-to-one correspondence between 
the ${\mathbb C}P^{N-1}$ moduli of 
the sine-Gordon soliton and 
the non-Abelian vortex.
The flux matching in Eq.~(\ref{eq:flux-matching}) 
also shows the coincidence of the topological charges of them:
\beq
 T_{\rm vortex} &=& \int d^2 x \tr F_{12}  
= 2 \pi k . \label{eq:charge}
\eeq
These precisely proves the identification of 
a non-Abelian vortex and a non-Abelian sine-Gordon kink 
inside the non-Abelian domain wall.

The $U(N)$ chiral Lagrangian in Eq.~(\ref{eq:eff}) 
with the quadratic pion mass term in Eq.~(\ref{eq:eff-Josephson2}) 
also admits non-Abelian sine-Gordon solitons 
\cite{Nitta:2014rxa}.
In this case, the modified sine-Gordon soliton
\beq
 u(x)
 = e^{i\theta_{\rm SG,2}(x^1)}
 = \exp (2 i \arctan \exp{\sqrt 2 m_2''  (x^1- X)} ),
\quad m_2''^2 \equiv {\pi \gamma_2 \over v^2}
\eeq
is embedded into $U$ in Eq.~(\ref{eq:embedding}).
Note that the range of $\theta$ for a single soliton is half the 
conventional sine-Gordon soliton. 
In the total configuration 
the conventional sine-Gordon soliton 
$\theta_{\rm SG}$ in Eq.~(\ref{eq:total}) 
is replaced by $\theta_{\rm SG,2}$ given here.
From the same calculation in Eq.~(\ref{eq:flux-matching}), 
we obtain
\beq
\int d^2 x F_{12} 
&=& 
V
{\rm diag} \left(
[\theta_{\rm SG,2} ]_{(x^1,x^2)=(-\infty,+\infty)}^{(x^1,x^2)=(+\infty,+\infty)}  
,0,\cdots,0 \right) V^\dagger\non
&=& V {\rm diag} (\pi, 0,\cdots,0) V^\dagger, \label{eq:flux-matching2}
\eeq
for the single soliton.
We thus have found that this solution carries a half color flux of 
a single non-Abelian vortex in the bulk. 
Therefore, one non-Abelian vortex must be split into 
two fractional non-Abelian fluxes when absorbed into 
the domain wall.
This is a non-Abelian extension of Ref.~\cite{Auzzi:2006ju}.

In the absence of any Josephson term, 
non-Abelian fluxes are diluted to infinity 
when absorbed into the domain wall, 
since the size of non-Abelian sine-Gordon soliton 
$m'^{-1}$ or $m_2'^{-1}$ goes to infinity 
in the limit of $\gamma$ to zero.

\subsection{3+1 dimensional configurations}
\begin{figure}[ht]
\begin{center}
\includegraphics[width=0.5\linewidth,keepaspectratio]{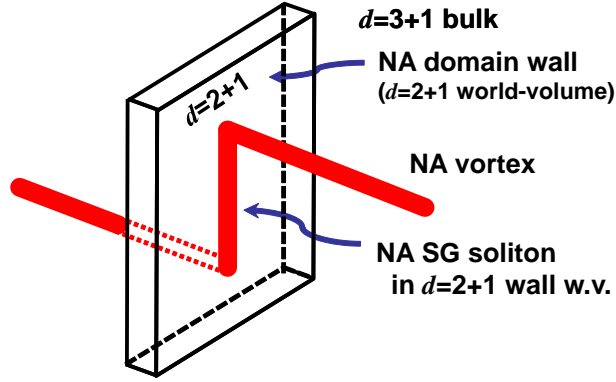}
\caption{
Three dimensional configurations of vortices on 
Josephson junctions 
with a linear Josephson term.
A non-Abelian vortex 
ends on a non-Abelian domain wall, 
becomes a non-Abelian sine-Gordon soliton 
inside the domain wall, 
and escapes to the other side of domain wall.
\label{fig:3d}
}
\end{center}
\end{figure}

We have been discussing configurations 
in dimension $d=2+1$.
In $d=3+1$, a non-Abelian vortex 
can end on a non-Abelian domain wall 
when they are perpendicular to each other \cite{Shifman:2003uh},
which is a non-Abelian generalization of a D-brane soliton
\cite{Gauntlett:2000de}.
In the presence of the conventional non-Abelian Josephson term 
in Eq.~(\ref{eq:Josephson}), 
a non-Abelian flux ending on a non-Abelian domain wall 
turns to a non-Abelian sine-Gordon soliton inside the wall, 
and it can escape to the other side of the domain wall, 
as illustrated in Fig.~\ref{fig:3d}.

%%%%%%%%%%%%%%%%%%%%%%
%%%%%%%%%%%%%%%%%%%%%%
%%%%%%%%%%%%%%%%%%%%%%
\begin{figure}[ht]
\begin{center}
\begin{tabular}{cc}
\includegraphics[width=0.44\linewidth,keepaspectratio]{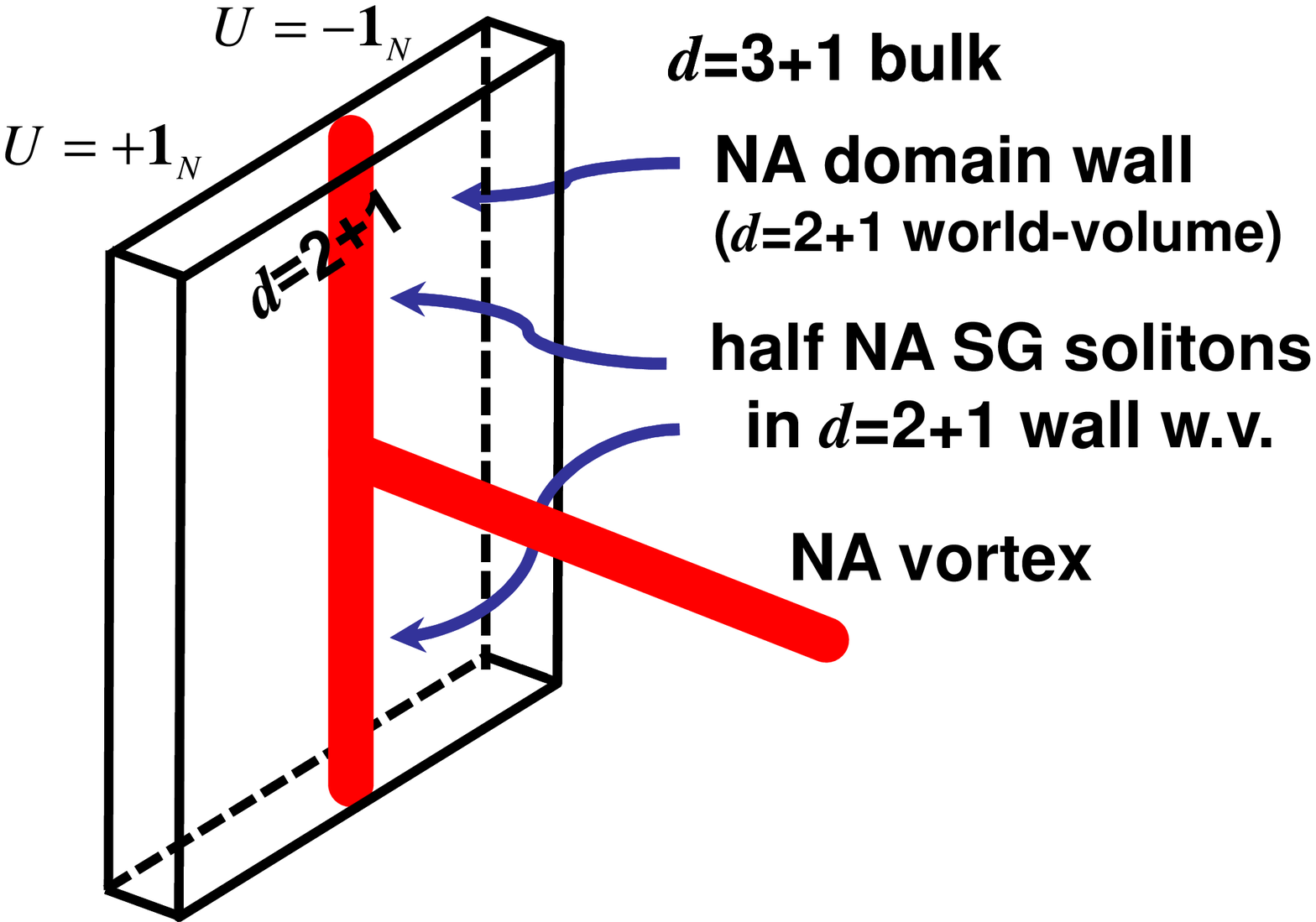}
&
\includegraphics[width=0.5\linewidth,keepaspectratio]{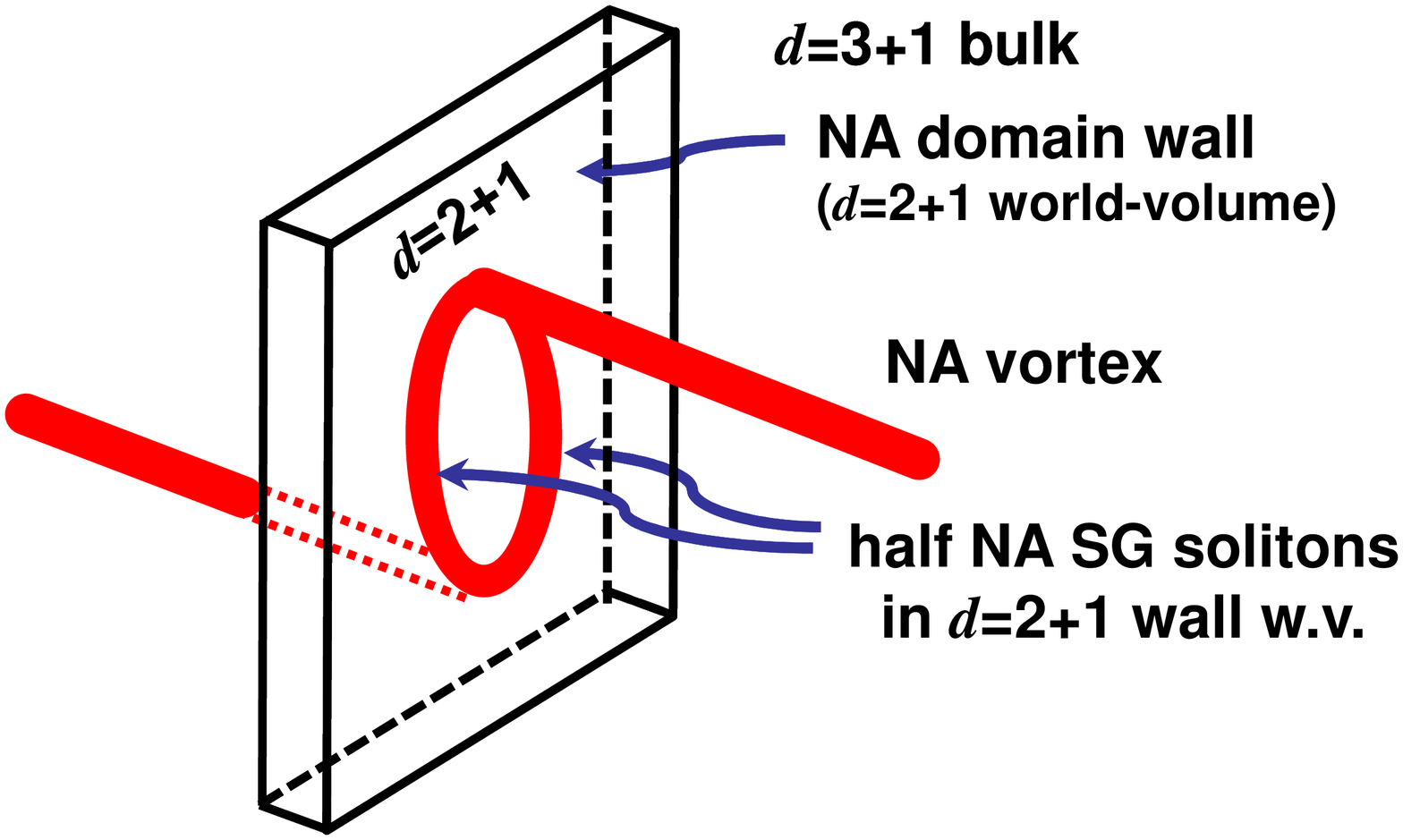}\\
(a) & (b) 
\end{tabular}
\caption{
Three dimensional configurations of vortices on 
Josephson junctions 
with a quadratic Josephson term.
(a) A non-Abelian vortex 
ending on a non-Abelian domain wall 
splits into two fractional fluxes inside the domain wall, 
represented as 
half non-Abelian sine-Gordon solitons. 
(b) They join to escape to the other side of domain wall, 
leaving a domain wall ring.
\label{fig:3d-2}
}
\end{center}
\end{figure}
%%%%%%%%%%%%%%%%%%%%%%

On the other hand, 
in the presence of the quadratic non-Abelian Josephson term 
in Eq.~(\ref{eq:Josephson2}), 
a non-Abelian flux ending on a non-Abelian domain wall 
is split into two fractional non-Abelian sine-Gordon solitons 
that separate the two vacua $U=\pm 1$,
as illustrated in Fig.~\ref{fig:3d-2}(a).
They have to join  
when they escape to the other side of the domain wall,  
as illustrated in Fig.~\ref{fig:3d-2}(b).
Consequently, there appears a domain wall ring 
that separates two vacua in $U=\pm {\bf 1}_N$ 
with two vortices on it.
This structure in $d=2+1$ is in fact known 
for the Abelian case in chiral p-wave superconductors 
\cite{Garaud:2012}.

In either case, when vortices do not end on both sides of 
the domain wall, the domain wall is logarithmically bent.
When two vortices end on the domain wall from the both sides 
there is a linear confinement force 
coming from the sine-Gordon soliton(s) between the two endpoints. 
Consequently, 
the vortices on the both sides tend to join at the same endpoint, 
if they are orthogonal to the domain wall.
If the vortices and the domain wall have an angle, 
the two endpoints will be separated and a vortex kink is formed, 
as is known for the Abelian Josephson junctions 
\cite{Blatter:1994}.

%%%%%%%%%%%%%%%%%%%%%
\section{Summary and Discussion \label{sec:summary} }

In this paper, we have constructed the non-Abelian Josephson 
vortex in a (flexible) non-Abelian Josephson junction 
of two non-Abelian color superconductors.
The effective field theory of a non-Abelian domain wall 
is the $U(N)$ chiral Lagrangian that  
has the conventional (modified or quadratic) pion mass term 
when the linear (quadratic) non-Abelian Josephson 
term is introduced in the model.
Then, we have shown that  
a non-Abelian sine-Gordon soliton found in Ref.~\cite{Nitta:2014rxa}
carries a (half) non-Abelian magnetic flux 
of a non-Abelian vortex in the bulk for 
the linear (quadratic) Josephson term,
and the ${\mathbb C}P^{N-1}$ moduli.
We thus have found that 
a non-Abelian vortex becomes 
one (two) non-Abelian sine-Gordon soliton(s)
when absorbed into the non-Abelian 
Josephson junction.
We have also discussed 3+1 dimensional configurations 
of Josephson vortices, 
that constitute a non-Abelian extension of a D-brane soliton.

Monopoles and Yang-Mills instantons trapped in the non-Abelian Josephson junction have been studied in Ref.~\cite{Nitta:2015mxa}. They are expressed as global vortices and Skyrmions, respectively, in the domain wall effective theory.

The $U(N)$ chiral Lagrangian appearing as 
the low-energy theory of QCD 
receives an axial anomaly for the $U(1)$ part 
and has a potential term along the $U(1)$ direction.
In this case,  there appear 
multiple domain walls 
other than the non-Abelian sine-Gordon soliton
\cite{Eto:2013bxa}.
On the other hand,  
the $U(N)$ chiral Lagrangian arising in the non-Abelian domain wall 
discussed in this paper 
does not have such a term.

In this paper, we have  considered $U(N)$ gauge theory 
but $SU(N)$ gauge group does not change the main results.
We still have a flux matching between 
a non-Abelian sine-Gordon soliton inside the non-Abelian domain wall 
and a non-Abelian vortex in the bulk.
Therefore, the results in this paper 
can be applied to color superconductors 
of high density quark matter. 
If quark matter is separated by an insulator 
for instance by some modulation such as crystalline superconductivity, 
it will give non-Abelian Josephson junctions.
Non-Abelian vortices there become non-Abelian Josephson 
vortices by trapped to insulating regions.

It was already pointed out in Ref.~\cite{Nitta:2014rxa} 
that the principal chiral model with 
arbitrary groups $G$ in the form of 
${G \times U(1) \over {\mathbb Z}_r}$ 
admits non-Abelian $G$ sine-Gordon solitons.
Non-Abelian vortices with this type of gauge groups 
were also found before \cite{Eto:2008yi}, 
such as $SO(N)$ and $USp(2N)$ groups \cite{Eto:2009bg}.
Therefore, by changing $U(N)$ groups in our model 
to ${G \times U(1) \over {\mathbb Z}_r}$, 
there should exists a non-Abelian domain wall 
whose effective theory is the principal chiral model with 
${G \times U(1) \over {\mathbb Z}_r}$, and  
non-Abelian vortices 
will become 
non-Abelian $G$ Josephson vortices 
as non-Abelian $G$ sine-Gordon solitons 
in the domain wall effective theory, 
if they are absorbed into the domain wall.

One may construct 
a Josephson junction of three superconductors 
that meet at one point, 
where the insulator is of a Y shape. 
As for a flexible version of this three junction,
we can use a domain wall junction solution 
for which exact solution is available \cite{Eto:2005cp}.

\section*{Acknowledgements}

This work is supported in part by 
Grant-in-Aid for Scientific Research (No. 23740198) 
and by the ``Topological Quantum Phenomena'' 
Grant-in-Aid for Scientific Research 
on Innovative Areas (No. 23103515)  
from the Ministry of Education, Culture, Sports, Science and Technology 
(MEXT) of Japan.

%%%%%%%%%%%%%%%%%%%%%%%%%%%%%%%%%%%%%%%%%%%%%%%%%%%%%%%%%%%%
%\newpage

%%%%%%%%%% References %%%%%%%%%%%%%%%%%%%%%%%%%
\newcommand{\J}[4]{{\sl #1} {\bf #2} (#3) #4}
\newcommand{\andJ}[3]{{\bf #1} (#2) #3}
\newcommand{\AP}{Ann.\ Phys.\ (N.Y.)}
\newcommand{\MPL}{Mod.\ Phys.\ Lett.}
\newcommand{\NP}{Nucl.\ Phys.}
\newcommand{\PL}{Phys.\ Lett.}
\newcommand{\PR}{ Phys.\ Rev.}
\newcommand{\PRL}{Phys.\ Rev.\ Lett.}
\newcommand{\PTP}{Prog.\ Theor.\ Phys.}
\newcommand{\hep}[1]{{\tt hep-th/{#1}}}
%%%%%%%%%%%%%%%%%%%%%%%%%%%%%%%%%%%%%%%%%%%%%%%


\begin{thebibliography}{100}
\bibitem{footnote1}
Exactly speaking, there are $N-2$ more vacua between these two vacua 
in the supersymmetric theory. Those are not relevant in our context.
\bibitem{footnote2}
In the supersymmetric extension, 
the full moduli space is 
the cotangent bundle over $U(N)$, 
$T^* U(N)$ larger than that in Eq.~(\ref{eq:wall-moduli}).
\end{thebibliography}
\end{document}